# Anomalous thermal expansion in a CuAl$_2$-type superconductor CoZr$_2$

Yoshikazu Mizuguchi[1]*, Md. Riad Kasem[1], and Yoichi Ikeda[2]

[1] *Department of Physics, Tokyo Metropolitan University, Hachioji, Tokyo 192-0397, Japan*
[2] *Institute for Materials Research, Tohoku University, Sendai, Miyagi 987-6543, Japan*

We investigated the anomalous thermal expansion in the CuAl$_2$-type superconductor CoZr$_2$ and the alloyed compounds of $Tr$Zr$_2$ ($Tr$: transition metal) over a wide temperature range. We performed neutron powder diffraction and X-ray powder diffraction on CoZr$_2$ and observed remarkably anisotropic thermal expansion with a $c$-axis negative thermal expansion constant of $\alpha_c < -15$ μK$^{-1}$ over a wide temperature range of $T$ = 50–573 K. With decreasing temperature, lattice constant $a$ decreased, whereas lattice constant $c$ continuously increased in CoZr$_2$. The cause of the anisotropic shrinkage/expansion of the $a$-axis/$c$-axis during cooling was explained by the small change in the Co-Zr bond and the systematic decrease in the Zr-Co-Zr angle. Similar thermal expansion was observed in the alloyed systems, (Fe,Co,Ni)Zr$_2$ and (Fe,Co,Ni,Rh,Ir)Zr$_2$, which suggests that this phenomenon is a common feature in the $Tr$Zr$_2$ system. We propose that zero-thermal expansion metals can be achieved by optimizing the contrasting thermal expansion of the $a$- and $c$-axes in $Tr$Zr$_2$.

Materials that exhibit negative thermal expansion (NTE) or zero-thermal expansion (ZTE) have been developed for a wide range of materials because they can be used for various applications, including precision instruments [1–5]. A typical NTE material is ZrW$_2$O$_8$, which exhibits a large isotropic NTE over a wide temperature ($T$) range [4]. In ZrW$_2$O$_8$, the cause of NTE is the flexible atomic bonds and linkage of units. In addition, NTE has been observed in various compounds that exhibit phase transition. The driving force includes a metal-insulator transition [5,6], magnetovolume effects [7,8], intermetallic charge transfer [9,10], valence crossover (or transition) [11,12], and ferroelectric transition [13,14]. Furthermore, NTE has been observed in various superconductors and related antiferromagnetic phases [15–21]. In most cases of NTE of superconductors, the cause of the NTE is linked to the emergence of the superconducting order parameter; hence, NTE is typically observed below the transition temperature ($T_c$) [3]. In addition, some superconductors exhibit NTE at temperatures above $T_c$; a typical example is the BSCCO superconductor [17]. Since the thermal expansion characteristics of superconductors are critical when fabricating superconducting wires or films, studies on the thermal expansion of superconductors have been performed over a wide temperature range [22–24]. Herein, we show anomalous anisotropic thermal expansion in $CoZr_2$ and alloyed $TrZr_2$ (Tr: Fe, Co, Ni, Rh, and Ir). We observed contrasting thermal expansions of the $a$- and $c$-axes in the tetragonal unit cell. In the wide temperature range of 50–572 K ($T$ = 572 K is the highest temperature examined in this study), $CoZr_2$ exhibited continuous positive thermal expansion (PTE) along the $a$-axis and NTE along the $c$-axis. As a result of the contrasting expansions, the lattice volume did not remarkably decrease with a decrease in temperature.

Figure 1 shows schematics of the crystal structure of $CoZr_2$ with a $CuAl_2$-type tetragonal structure (space group #140). The framework was composed of $CoZr_8$ polyhedron units stacked along the $c$-axis. $T_c$ of $TrZr_2$ with $Tr$ = Ni, Co, Rh, and Ir were 3.2, 5.2, 11.3, and 7.5 K, respectively [25]. The electronic density of states (DOS) near the Fermi energy ($E_F$) was mainly due to the hybridization of Zr-$d$ and Co-$d$ orbitals [26]. Noticeably, the $T_c$ of $CoZr_2$ increased with the application of external pressures: $T_c$ = 9.5 K at 8 GPa [26]. The large pressure effect implies a flexible crystal structure, which would significantly modify the electronic DOS and/or electron-phonon coupling in the system under high pressures. In addition, we recently developed alloyed compounds of $TrZr_2$ using the high-entropy alloy (HEA) concept [27–29]. Interestingly, the $T_c$ of $TrZr_2$ increased with increasing lattice constant $c$, whereas the variation in the element, concentration, and configurational entropy of mixing at the $Tr$ site did not have an impact on $T_c$, although alloying at the $Tr$ site introduced a strong disorder. This trend also implies the importance of the chemical bonds along the $c$-axis in $TrZr_2$ superconductors. Based on the relationship between superconductivity and the crystal structure of $TrZr_2$, we performed crystal structure analyses for $CoZr_2$ and alloyed compounds over a wide temperature range.

Polycrystalline samples of $CoZr_2$, $Fe_{1/3}Co_{1/3}Ni_{1/3}Zr_2$ (($Fe,Co,Ni)Zr_2$), and $Fe_{0.2}Co_{0.2}Ni_{0.2}Rh_{0.2}Ir_{0.2}Zr_2$ (($Fe,Co,Ni,Rh,Ir)Zr_2$) were prepared by melting Fe (99.9%), Co (99%), Ni (99.9%), Rh (99.9%), and Ir (99.9%) powders and Zr (99.2 %) plates using an arc furnace, as described in Ref. 29. The phase purity and crystal structure were examined using X-ray powder diffraction (XRD) with Cu-K$\alpha$ radiation using the $\theta$–$2\theta$ method on a Miniflex 600 (Rigaku) equipped with a high-resolution semiconductor detector D/tex-Ultra. For the high-temperature XRD on a Miniflex-600, the sample temperature was controlled using a BTS 500 attachment. For low-temperature experiments, we performed neutron powder diffraction (NPD) using a HERMES diffractometer [30] installed at the T1-3 guide port of the JRR-3 of the Japan Atomic Energy Agency, Tokai. A thermal neutron beam was monochromatized to 2.197 (1) Å using a vertically-focused Ge (331) monochromator. Typical instrumental parameters were determined by analyzing the line positions and line shapes of a standard reference material (LaB$_6$, NIST 660c) [31]. Powder samples were sealed in a vanadium cylinder cell with a diameter of 6 mm (0.1 mm thick) and a length of approximately 60 mm in a $^4$He gas atmosphere. A closed-cycle refrigerator was used to cool the samples from the base temperature ($T \sim 7$ K) to room temperature. Notably, the samples used in the NPD and XRD were different batches. The $T_c$ of the samples was confirmed by magnetization measurements using a superconducting interference device (SQUID) with an applied field of 10 Oe on an MPMS3 (Magnetic Property Measurement System 3, Quantum Design); the magnetization data are shown in Supplementary Data (Fig. S1) [32]. All the examined samples showed bulk superconductivity with a large diamagnetic signal, and the measured $T_c$ values for CoZr$_2$, (Fe,Co,Ni)Zr$_2$, and (Fe,Co,Ni,Rh,Ir)Zr$_2$ were 6.0, 3.3, and 5.4 K, respectively. The obtained XRD and NPD patterns were refined using the Rietveld method using RIETAN-FP [33], and schematics of the refined crystal structure were depicted using VESTA [34]. In the NPD and XRD patterns, we normalized the data by the highest intensity to compare the shift in the peaks with changing temperatures. The Rietveld refinement results are represented as the mean values with standard errors. The number of data points for the linear fitting of the lattice constants is shown in the plots in Figs. 2 and 3.

Figure 2(a) shows the NPD patterns of CoZr$_2$ obtained at $T = 7$, 50, 100, 170, and 293 K (room temperature). The profiles at $T = 7$ K and 293 K were similar, and no crystal structural transitions or magnetic ordering were observed. As shown in Fig. 2(b), we observed a decrease in lattice constant $a$ (from the 200 peak) and an increase in lattice constant $c$ (from the 002 peak). To determine the temperature dependences of the lattice constants, the NPD patterns were analyzed using the Rietveld method with a three-phase mode, as shown in Figs. 2(c) and S2 (Supplementary Data). The refined structural parameters are summarized in Table I. Here, we identified two impurity phases of CoZr$_3$ (16 wt%, space group #63) and Co-Zr (5 wt%, space group #225). We analyzed the isotropic displacement parameters ($U_{iso}$) at all temperatures for the Co and Zr sites. Since the obtained $U_{iso}$ (Co) values at $T = 7$ K and 50 K were approximately zero within the errors of the fitting quality of the data, we performed refinements by fixing $U_{iso}$ (Co) as $U_{iso}$ (Zr) for the data at $T = 7$ K and 50 K. Figures 2(d–f) show the temperature dependences of lattice constants $a$, $c$, and $V$ for CoZr$_2$. With decreasing temperature, lattice constant $a$ decreased, which was a PTE, whereas lattice constant $c$ increased, which was an NTE. The results of the high-temperature

XRD also exhibited the same trend (see the Supplementary Data for the XRD patterns). Since the *a*- and *c*-axes exhibited contrasting PTE/NTE, the lattice volume (*V*) did not show monotonous changes. These results imply that the *Tr*Zr$_2$ system is a potential ZTE material after tuning the PTE and NTE of the *a*- and *c*-axes. For the *c*-axis NTE, the thermal expansion constant ($\alpha_c$) was estimated to be $-15$ μK$^{-1}$ for low-temperature NPD data and $-28$ μK$^{-1}$ for high-temperature XRD data using the formula $\alpha_c = [1/c\,(300\,\text{K})]/(\text{d}c/\text{d}T)$. The reported value of the linear thermal expansion constant ($\alpha$) for typical NTE materials are $\alpha = -9$ μK$^{-1}$ for ZrW$_2$O$_8$ [4], $-26$ μK$^{-1}$ for LaFe$_{10.5}$Co$_{1.0}$Si$_{1.5}$ [7], $-0.7$ μK$^{-1}$ for Ba(Fe$_{0.962}$Co$_{0.038}$)$_2$As$_2$ [15], $-0.8$ μK$^{-1}$ for PrFeAsO [35], and $-87$ μK$^{-1}$ for BSCCO [17]. The $\alpha_c$ observed in CoZr$_2$ was comparable to that of LaFe$_{10.5}$Co$_{1.0}$Si$_{1.5}$, which was relatively large among the NTE of superconductors. The linear thermal expansion constant along the *a*-axis ($\alpha_a$) was also estimated, as shown in Fig. 2(a).

To understand the cause of the anomalous (or anisotropic) thermal expansion of the *a*- and *c*-axes, the atomic distances of Co-Zr, Co-Co, and Zr-Zr were determined from the NPD data and are summarized in Table I (see Fig. 1 for the labels of the distances and the angle). The Co-Co (i) distance was directly linked to lattice constant *c*, which increased with decreasing temperature. The Co-Co (ii) and Zr-Zr (ii) distances decreased with decreasing temperature, owing to the *a*-axis compression. The Co-Zr and Zr-Zr (i) distances did not change significantly at low temperatures. However, the Zr-Co-Zr angle exhibited a systematic decrease as the temperature decreased. The temperature evolutions of the Co-Zr and Zr-Zr (i) distances and Zr-Co-Zr angle indicate that the shape of the CoZr$_8$ polyhedrons was flexibly modified during cooling. Therefore, we conclude that the anomalous anisotropic thermal expansion (elongation of the unit cell) in CoZr$_2$ was induced by the robust Co-Zr and Zr-Zr (i) distances to the temperature change and the flexible Zr-Co-Zr angle in the CoZr$_8$ polyhedron. As reviewed in Ref. 36, the NTE in ReO$_3$-type materials is caused by the flexibility of the structure, and the cause of the flexible bonds is discussed in terms of the anisotropic atomic vibration (displacement). To examine anisotropic atomic displacement in CoZr$_2$, we performed anisotropic analyses of atomic displacement parameters $U_{11}$ and $U_{33}$ along the *a* and *c*-axes, respectively. Since the anisotropic analysis of the Co site was unsuccessful at low temperatures ($T \leq 170$ K), only the results for $T = 293$ K are shown; for $T = 170$ K, $U_{33}$ (Co) was approximately zero after the refinement. The estimated values were $U_{11}$ (Co) = 0.052(6) Å$^2$, $U_{33}$ (Co) = 0.013(6) Å$^2$, $U_{11}$ (Zr) = 0.023(2) Å$^2$, and $U_{33}$ (Zr) = 0.014(2) Å$^2$. $U_{11}$ (Co) was larger than the other parameters, which could be linked to the *c*-axis NTE. To understand the cause, local structural analyses are required;

extended X-ray absorption fine structure or pair density function analyses can provide further information on the cause of NTE in $CoZr_2$. Similar contrasting axes thermal expansion have also been observed in cordierite ($Mg_2Al_4Si_5O_{18}$) and was well explained through ab initio molecular dynamics (MD) simulation [37]. Therefore, MD simulations should be useful for understanding the cause of NTE along the $c$-axis of $CoZr_2$.

In this study, we examined the low-temperature crystal structures of the alloyed compounds $(Fe,Co,Ni)Zr_2$ and $(Fe,Co,Ni,Rh,Ir)Zr_2$. For $(Fe,Co,Ni)Zr_2$, a three-phase analysis similar to that for $CoZr_2$ was performed. For $(Fe,Co,Ni,Rh,Ir)Zr_2$, owing to the presence of unknown impurity peaks, we performed a single-phase analysis. See the Supplementary Data for the Rietveld refinement results of these samples. Furthermore, high-temperature XRD was performed to approximate the lattice constants at high temperatures using samples obtained from different batches (see Fig. S5 in the Supplementary Data). Figures 3(a–f) show the temperature dependences of lattice constants $a$, $c$, and $V$ for the alloyed samples. Noticeably, contrasting thermal expansion, PTE along the $a$-axis and NTE along the $c$-axis were observed for both $(Fe,Co,Ni)Zr_2$ and $(Fe,Co,Ni,Rh,Ir)Zr_2$. In addition, the trend of the temperature evolution of the $TrZr_8$ polyhedron was similar to that for $CoZr_2$, that is, the change in the $Tr$-$Zr$ distance was small, and the $Zr$-$Tr$-$Zr$ angle systematically decreased as the temperature decreased (see Tables in the Supplementary Data for the refined parameters).

We briefly discuss the cause of the anomalous (contrasting anisotropic) thermal expansion in $TrZr_2$. Based on the binary Co-Zr phase diagram, the $CuAl_2$-type $CoZr_2$ phase exhibited congruent melting, and no structural transition was expected at temperatures higher than 572 K. In addition, no magnetic ordering was observed in the NPD results. Furthermore, the NTE of the $c$-axis occurred at temperatures above a $T_c$ value of 6.0 K for $CoZr_2$. Although we have to confirm a possible change in the valence state of $Tr$ (i.e., electronic and bonding states of $Tr$) to exclude the possibility of valence-state-driven NTE, we propose that the robust $Tr$-$Zr$ bonds and the flexibility of the bond angle in the $TrZr_8$ polyhedron are the reason of the anomalous anisotropic thermal expansion in $TrZr_2$.

Since the $a$- and $c$-axes of $TrZr_2$ show contrasting thermal expansion (PTE and NTE), one of the expected applications is ZTE by tuning the changes in both axes. In fact, the examined samples slightly expand upon cooling. The volume compression rates calculated using $[V(293\ \text{K})-V(50\ \text{K})]/V(293\ \text{K})$ for $CoZr_2$, $(Fe,Co,Ni)Zr_2$, and $(Fe,Co,Ni,Rh,Ir)Zr_2$ were −0.55%, −0.46%, and −0.11%, respectively. Here, we used the data at 50 K because the NTE of the $c$-axis was observed down to 50 K for $CoZr_2$. This trend suggests that alloying the $Tr$ site using the HEA concept is useful for achieving ZTE (volume ZTE) in $Tr$Zr$_2$. Furthermore, tunable volume thermal expansion is important in the manufacture of superconductors and superconducting devices [22–24]. Comparing the estimated linear thermal expansion constants, $\alpha_a$ and $\alpha_c$, we observed that the $a$-axis thermal expansion below room temperature was affected and suppressed by alloying (with increasing configurational entropy of mixing at the $Tr$ site), whereas the $c$-axis thermal expansion was not. Chemical substitution introduces local disorder in materials and modifies thermal expansion [36]. Therefore, tuning the configurational entropy of mixing is useful for achieving ZTE. Although we were not able to determine the cause of the suppression of the $a$-axis thermal expansion, the large isotropic displacement parameter ($U_{iso}$) for the Zr site for the alloyed systems, which is shown in Fig. 4, could be attributed to the weakened $a$-axis thermal expansion due to alloying. The weakened temperature dependence of $U_{iso}$ indicates the local inhomogeneity of the position of Zr atoms and the atomic bonds.

Another prospect is the enhancement of $T_c$ in $Tr$Zr$_2$. As discussed in Ref. 29, a larger lattice constant $c$ is preferable for a higher $T_c$ in $Tr$Zr$_2$. Therefore, if we could clarify the cause of the NTE of the $c$-axis in $Tr$Zr$_2$, a new superconductor with a $T_c$ exceeding 12 K, which is higher than the $T_c$ of RhZr$_2$, could be designed. In addition, crystal structure analyses under high pressure are required to verify the cause of the enhancement of $T_c$ due to pressure and the link to the anomalous thermal expansion in $Tr$Zr$_2$ reported here.

In conclusion, we analyzed the crystal structures of the CuAl$_2$-type transition-metal zirconides CoZr$_2$, (Fe,Co,Ni)Zr$_2$, and (Fe,Co,Ni,Rh,Ir)Zr$_2$ using NPD at low temperatures and XRD at high temperatures. With decreasing temperatures, all the samples exhibited $a$-axis shrinkage (PTE) and $c$-axis expansion (NTE). The anomalous (contrasting and anisotropic) thermal expansion of the $a$- and $c$-axes in $Tr$Zr$_2$ was explained by the small change in the $Tr$-Zr distance and the systematic decrease in the Zr-$Tr$-Zr angle, which resulted in the elongation of the unit cell along the $c$-axis. Although the cause of the structural change remains unclear, we consider that the flexible bonding of the $Tr$Zr$_8$ polyhedron units is essential for the NTE of the $c$-axis in $Tr$Zr$_2$.

**Acknowledgment**

The authors would like to thank M. Fujita and O. Miura for their support with the experiments and discussions. This study was performed under the GIMRT Program of the Institute for Materials Research, Tohoku University (CN: Center of Neutron Science for Advanced Materials: Proposal No. 202112-CNKXX-0001). This work was conducted using the JRR-3 program managed by the Institute for Solid State Physics, University of Tokyo (T1-3 HERMES IRT program: Proposal No. 22410). This work was partly supported by a Grant-in-Aid for Scientific Research (KAKENHI) (Proposal Nos. 21K18834, 21H00151, 19H05164, and 21H00139) and the Tokyo Government Advanced Research (H31-1).

*mizugu@tmu.ac.jp

78, 140503 (2008).

36) T. Tokizono, Y. Tsuru, T. Atsumi, N. Hosokawa, and T. Ohnuma, J. Cer. Soc. Jpn. 124, 744 (2016).

37) Q. Li, K. Lin, Z. Liu, L. Hu, Y. Cao, J. Chen, and X. Xing, Chem. Rev. 122, 8438 (2022).
9

Table I. Structural parameters of CoZr$_2$ obtained from the Rietveld refinement. The atomic coordinates for the Zr and Co site are ($x$, $x$+0.5, 0) and (0, 0, 0.25), respectively.

| | $T$ = 293 K | $T$ = 170 K | $T$ = 100 K | $T$ = 50 K | $T$ = 7 K |
|---|---|---|---|---|---|
| Space group | | | $I4/mcm$ (#140) | | |
| $a$ (Å) | 6.3584(3) | 6.3409(3) | 6.3337(3) | 6.3287(3) | 6.3274(3) |
| $c$ (Å) | 5.5002(3) | 5.5074(3) | 5.5159(2) | 5.5217(2) | 5.5211(3) |
| $V$ (Å$^3$) | 222.37(2) | 221.44(2) | 221.28(2) | 221.16(2) | 221.04(2) |
| $R_{wp}$ (%) | 16.41 | 16.02 | 14.31 | 13.84 | 14.8 |
| $S$ | 1.46 | 1.64 | 1.54 | 1.50 | 1.87 |
| $x$ (Zr) | 0.1710(2) | 0.1720(2) | 0.1717(2) | 0.1715(2) | 0.1721(2) |
| $U_{iso}$ (Co) (Å$^2$) | 0.034(3) | 0.07(2) | 0.08(2) | 0.0045 (fixed) | 0.0112 (fixed) |
| $U_{iso}$ (Zr) (Å$^2$) | 0.0258(13) | 0.0160(13) | 0.0110(10) | 0.0045(9) | 0.0112(11) |
| Co-Zr distance (Å) | 2.7293(12) | 2.7223(12) | 2.7218(11) | 2.7214(11) | 2.7195(11) |
| Co-Co (i) distance (Å) | 2.7501(2) | 2.7537(2) | 2.75795(11) | 2.76085(11) | 2.7606(2) |
| Co-Co (ii) distance (Å) | 4.4961(3) | 4.4837(3) | 4.4786(3) | 4.4751(3) | 4.4742(2) |
| Zr-Zr (i) distance (Å) | 3.0954(12) | 3.0887(12) | 3.0942(12) | 3.0979(12) | 3.0926(9) |
| Zr-Zr (ii) distance (Å) | 3.334(2) | 3.321(2) | 3.319(2) | 3.317(2) | 3.314(2) |
| Zr-Co-Zr angle (deg.) | 119.49(3) | 119.24(3) | 119.12(3) | 119.04(3) | 119.00(4) |

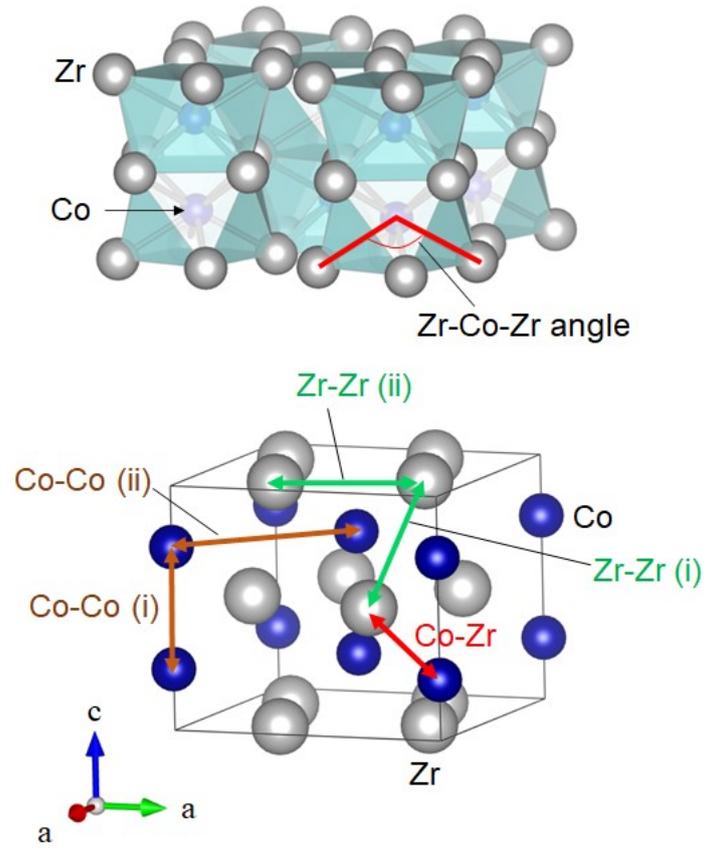

Fig.1. Schematic images of the crystal structure of CoZr$_2$. The solid line in the lower figure shows the unit cell of CoZr$_2$.

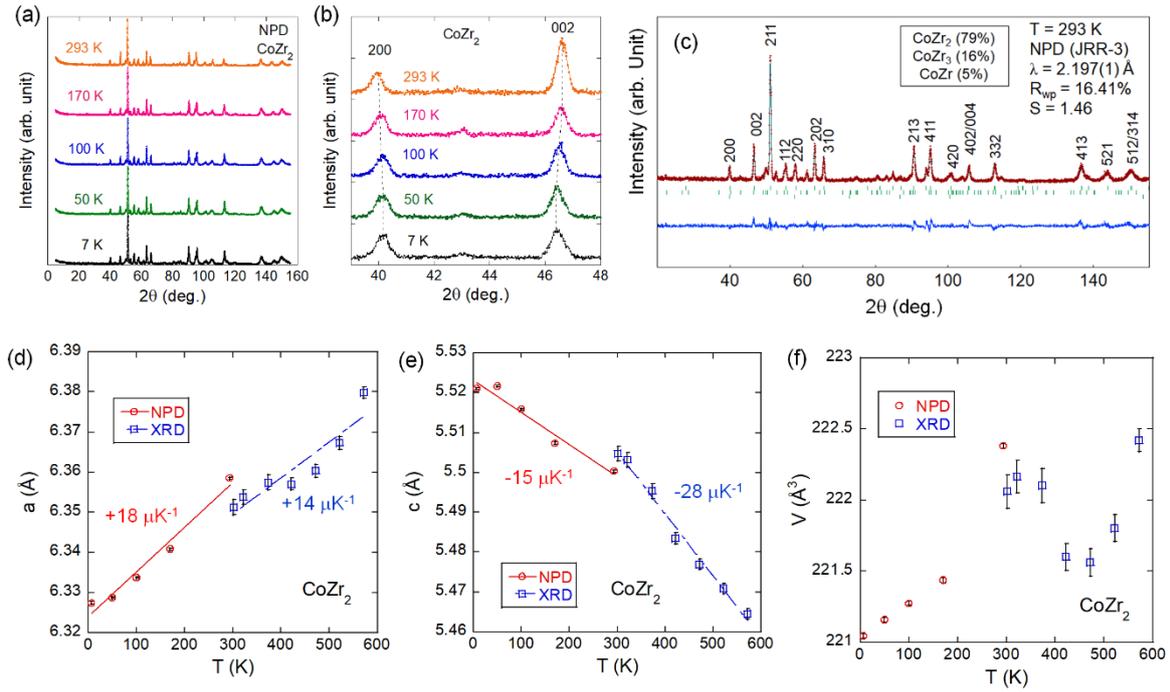

Fig. 2. (a) NPD patterns for CoZr$_2$ collected at $T$ = 7, 50, 100, 170, 293 K. (b) NPD patterns near the 200 and 002 peaks. (c) Rietveld refinement results for the NPD data for CoZr$_2$ ($T$ = 293 K). The numbers indicate Miller indices. (d–f) Temperature dependence of lattice constant $a$, $c$, and $V$ for CoZr$_2$. NPD and XRD denote neutron powder diffraction and X-ray powder diffraction, respectively. The error bars in (d,e,f) are a standard error determined by Rietveld refinement. The estimated linear thermal expansion constants are +18 ± 2 µK$^{-1}$ ($a$ by low-temperature NPD), +14 ± 3 µK$^{-1}$ ($a$ by high-temperature XRD), -15 ± 2 µK$^{-1}$ ($c$ by low-temperature NPD), and -28 ± 1 µK$^{-1}$ ($c$ by high-temperature XRD).

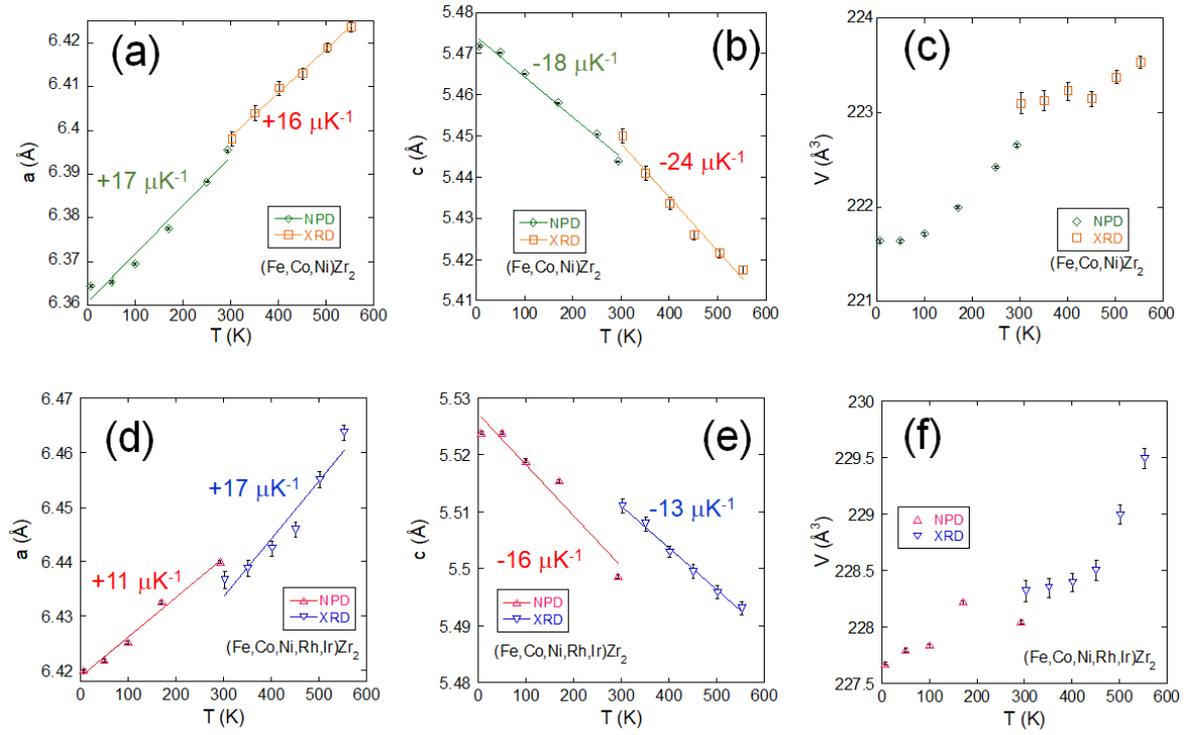

Fig. 3. Lattice constants determined from NPD and XRD for alloyed $Tr$Zr$_2$. The error bars are a standard error determined by Rietveld refinement. (a–c) Temperature dependence of lattice constant $a$, $c$, and $V$ for Fe$_{1/3}$Co$_{1/3}$Ni$_{1/3}$Zr$_2$. The estimated linear thermal expansion constants are +17 ± 2 µK$^{-1}$ ($a$ by low-temperature NPD), +16 ± 1 µK$^{-1}$ ($a$ by high-temperature XRD), -18 ± 1 µK$^{-1}$ ($c$ by low-temperature NPD), and -24 ± 2 µK$^{-1}$ ($c$ by high-temperature XRD). (d–f) Temperature dependence of lattice constant $a$, $c$, and $V$ for Fe$_{0.2}$Co$_{0.2}$Ni$_{0.2}$Rh$_{0.2}$Ir$_{0.2}$Zr$_2$. The estimated linear thermal expansion constants are +11 ± 1 µK$^{-1}$ ($a$ by low-temperature NPD), +17 ± 2 µK$^{-1}$ ($a$ by high-temperature XRD), -16 ± 2 µK$^{-1}$ ($c$ by low-temperature NPD), and -13 ± 1 µK$^{-1}$ ($c$ by high-temperature XRD).

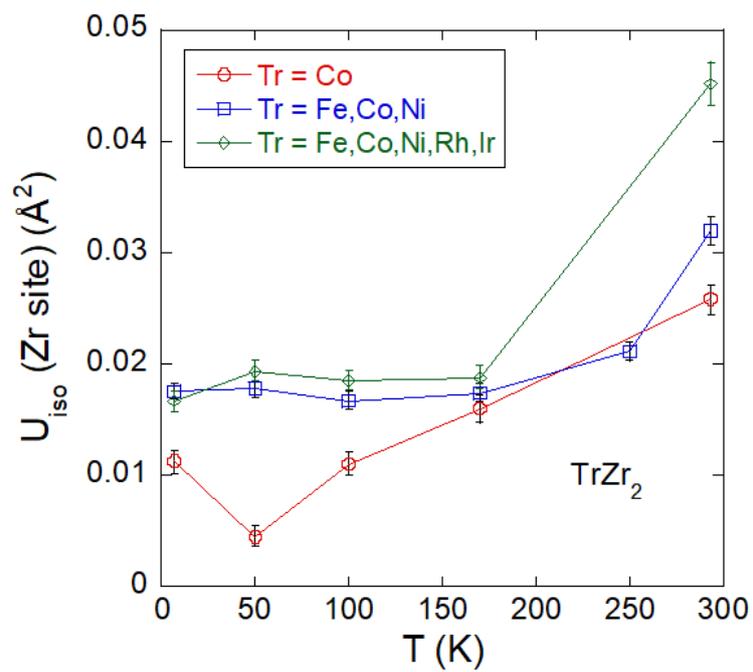

Fig. 4. Temperature dependences of isotropic displacement parameter ($U_{iso}$) for $CoZr_2$, $(Fe,Co,Ni)Zr_2$, and $(Fe,Co,Ni,Rh,Ir)Zr_2$.